\documentclass[prb,twocolumn,showpacs,floatfix]{revtex4}
\usepackage{epstopdf}
\usepackage{graphicx}
\usepackage{dcolumn}
\usepackage{bm}
\begin{document}
\title{Dynamical Response of Single Bi-layer Spin Model : A Theoretical Analysis}
\author{R. Chaudhury$^1$,  F. Demmel$^2$ and T. Chatterji$^3$ }
\address{$^1$ S.N. Bose National Centre for Basic Sciences, Block JD, Sector III, Salt Lake Kolkatta 700 098, India\\
$^2$ISIS Facility, Rutherford Appleton Lanoratory, Didcot OX11 0QX, United Kingdom\\
$^3$Institut Laue-Langevin, BP 156, 38042 Grenoble Cedex 9, France\\
}
\date{\today}
\begin{abstract}
 The spin dynamics of a single bi-layer ferromagnetic model, as proposed for a manganite system $La_{1.2}Sr_{1.8}Mn_{2}O_{7}$ as an approximate minimal model, is studied both below and above the bulk magnetic transition temperature using the semi-classical Monte Carlo-molecular dynamics technique. The quantities studied are:- (i) the static spin configurations, (ii) the spin auto-correlation function (SAF) $C(t)$ and (iii) the dynamical structure function (DSF) $S(\bf q, \omega)$. The major aim is to probe the nature of collective modes and in particular to ascertain whether the propagating modes can exist above the ordering temperature. We find that the typical spin configurations contain a high degree of mis-aligned spins, particularly in the high temperature phase. Nevertheless, in the low temperature phase a long range ferromagnetic ordering can be seen. The shape of the curves corresponding to  $S(\bf q, \omega)$ vs. $\omega$ in the constant $\bf q$-scan are found in general to be quite complex viz. containing peaks, cusps and broad plateau. Moreover, in a large regime of $(\bf q, \omega)$ space, $S(\bf q ,\omega)$ is found to be negative, signalling a total breakdown of the semi-classical approximation in the presence of the enormous thermodynamic and quantum fluctuations. Considering the physically allowed regime only, we can theoretically extract an effective $\omega$ vs. $q$ dispersion curve for the collective excitations, which exhibits a slope opposite to that expected for a full three-dimensional system and disagrees with the results of inelastic neutron scattering experiments in the nature of the dispersion curve along the $<100>$ direction in the ordered phase. The theoretical results also exhibit the existence of propagating modes even in the high temperature (disordered) phase from our calculations based on a minimal single bi-layer ferromagnetic model.
 \end{abstract}
\maketitle

\section{Introduction}
 The problem of understanding the spin dynamics of the spin models, both classical and quantum, at temperatures above the ordering temperature is a long standing challenging problem in condensed matter physics [1-4]. There are key questions regarding the nature of the collective modes viz. whether they are propagating or damped. This problem has assumed greater significance and importance in the context of the recent experimental studies on the various types of low-dimensional and layered magnetic systems like Manganites, Cuprates, Bromides etc [5-7]. All of these systems are modelled by either ferromagnetic or antiferromagnetic type of Heisenberg spin Hamiltonians with the presence of either Ising or XY-like anisotropy in a few cases. Furthermore, most of these systems have both intra-layer and comparatively weaker inter-layer exchange interactions and this makes the study of spin dynamics even more challenging. 

 In this short communication, we present our calculational results for the spin dynamics with parameters appropriate to those of $La_{1.2}Sr_{1.8}Mn_{2}O_{7}$, both above and below the bulk ferromagnetic ordering temperature. Detailed input from crystallographic and other measurements leads to the conclusion that in this system the intra-bilayer exchange interaction is much much larger than the inter-bilayer interaction. This motivates us to effectively model the magnetic properties of the system by a single bi-layer spin Hamiltonian. Furthermore, the intra-bilayer interaction has two components $viz.$ the intra-layer interaction represented by $J_1$ and the inter-layer one represented by $J_2$, where the magnitude of $J_1$ is larger than that of $J_2$ and both are ferromagnetic in nature. We carry out detailed Monte Carlo-molecular dynamics (MCMD) studies on the above model by considering a lattice of finite size with dimensions $50 \times 50 \times2$. This size then represents a single bi-layer system with size of each layer being $50 \times 50$. The calculations were performed in the low temperature phase ($16$K) as well as in the high temperature phase ($200$K). The experimental value for the Curie temperature ($T_c$) of this system is $128$K [3,4]. Various static and dynamic properties were studied. The important amongst them are :- (i) the static real space spin configurations, (ii) the dynamic structure function and (iii) the spin auto-correlation function. We try to present a detailed microscopic analysis of these results and compare them with the available experimental results. The experimental results available so far, are all based on the inelastic neutron scattering measurements carried out on the colossal magnetoresistive bilayer manganite $La_{1.2}Sr_{1.8}Mn_{2}O_{7}$. These have direct bearing on our theoretical results for the dynamic structure function $S({\bf q}, \omega)$. It would however be equally important and interesting to compare our results for the spin auto-correlation function with the corresponding results from Perturbed Angular Correlation (PAC) or ESR or Muon Spin Resonance experiments for the above system, when available. 

 The theoretical study of static spin configurations is also extremely challenging and gives us a lot of useful information regarding the intrinsic fluctuations originating from the spin model. These again in general, can have contributions from the quantum effects, the thermodynamic effects and also spatial dimensionality as well as the finite size effects. It is however very diffficult to experimentally determine the static spin configurations in real space directly. Nevertheless, the real space spin configurations do play a very crucial role in determining the nature of both the static and dynamic spin correlation functions. 
\section{ Model and brief calculational procedure}
The effective spin model we employ is essentially a nearest and next nearest neighbour quasi-two-dimensional (describing a single bi-layer) quantum Heisenberg spin Hamiltonian, given by [3]
\begin{equation}
{\mathcal H} = -[\Sigma_{<ij>} J_1{\bf S}_i\cdot{\bf S}_j + \Sigma_{<<ij>>}J_2{\bf S}_i\cdot{\bf S}_j]
\end{equation}
 where $J_1$ and $J_2$ are the nearest and the next nearest neighbour ferromagnetic exchange interactions respectively. $J_1$ operates within a layer whereas $J_2$ operates between the two layers of the bilayer. The above model corresponds to a single bi-layer, as explained earlier. It may further be mentioned that the magnetism arises from the electrons of $Mn$ atoms possessing an effective spin $S=1.8$ [3]. The values of $J_1$ and $J_2$ are $4.8$ mev and $1.7$ mev respectively [3,4].

 We carry out a semi-classical MCMD calculations on the above model. The details of this technique can be seen in the earlier works [1]. The semiclassical treatment is expected to hold for the present model, as the effective spin value per site is close to $2$.

 We implemented Metropolis algorithm for cooling, starting from infinite temperature. Corresponding to $T=16K$, we stored spin configurations with MC ages between $5000$ and $15,000$ MCS (MC steps per spin). For $200K$ however, we stored spin configurations with MC ages upto $30,000$ MCS.

 We carried out the tests for stabilization of various thermodynamic quantities  to ensure the attainment of thermal equilibrium. To our surprise however, we discovered that in our present model system even when the above stabilization or convergence sets in, the fluctuations in the static properties are unexpectedly large. We address this issue in the section on Results.

\section{ Results}
 In this section, we present our calculational results for both static and the dynamic properties. Moreover we highlight the salient features of our results.

 Let us first discuss about the static properties. We analysed the real space spin configurations and calculated the average magnetization, average energy and the specific heat from these configurations. This was done for both the temperatures, T = 16K and 200K. 

 We observed that in the low temperature phase ($16$K) the most probable spin configurations generated by MC, have a moderate degree of misalignment of spins, $viz.$ $17$ percent. The average magnetization per site is however almost constant. The ratio of the square root of the specific heat to the absolute value of the average energy is close to $0.4$, which is much larger than the expected magnitude for our system in the thermal equilibrium. This brings into focus the substantial and non-trivial contribution of the fluctuations to the static properties even in the low temperature phase.\\

\begin{figure}
\includegraphics[scale=.5,angle=0]{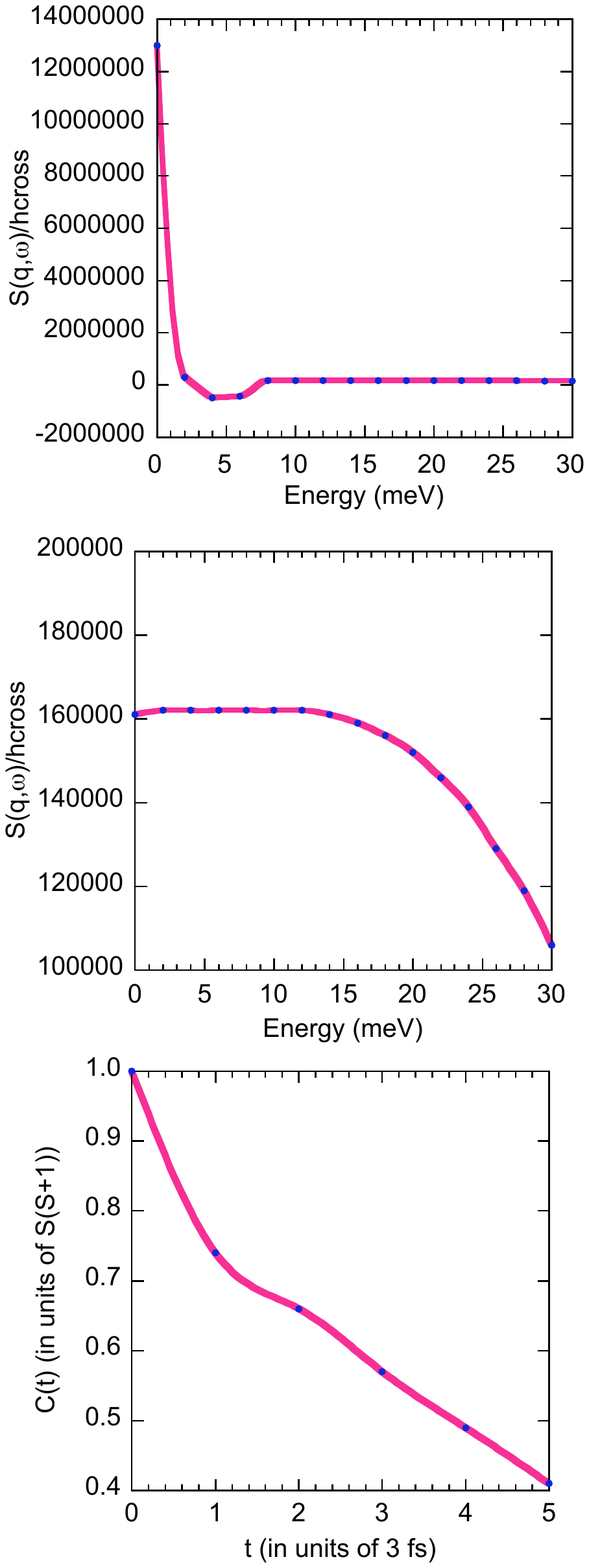}
\caption{ (Top and middle panel) $S({\bf q},\omega)$ vs. $\omega$ for $q=0.1$ rlu and $q=0.5$ rlu along $<100>$ direction at $T=16K$; (bottom panel) $C(t)$ vs. $t$ at $T=16K$.}
\label{fig1}
\end{figure}

\begin{figure}
\includegraphics[scale=.5,angle=0]{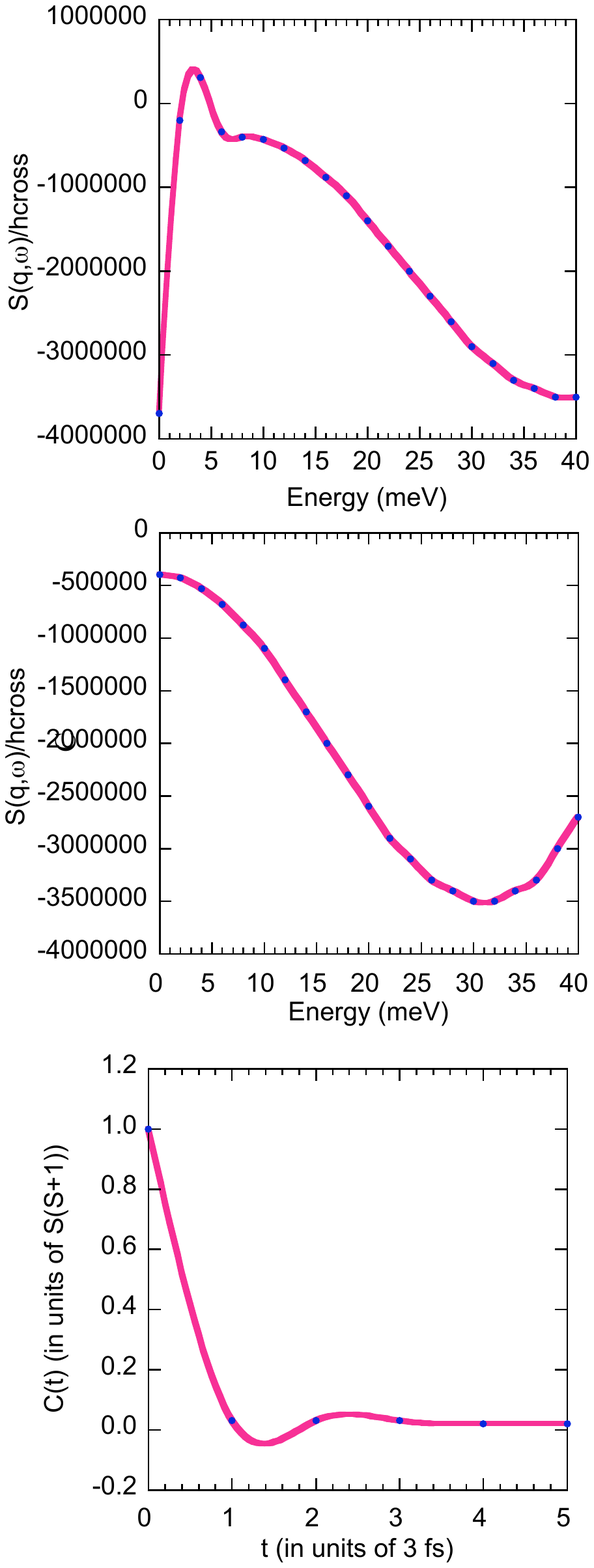}
\caption{(Top and middle panel) $S({\bf q},\omega)$ vs. $\omega$ for $q=0.1$ rlu and $q=0.5$ rlu along $<100>$ direction at $T=200K$; (bottom panel) $C(t)$ vs. $t$ at $T=200K$. }
\label{fig2}
\end{figure}

 In the high temperature regime $viz.$ at $200$K, our MC data show that the misalignment of spins in a typical spin configuration is around $80$ to $90$ percent. Moreover the average magnetization at a site also exhibits a misalignment (disorder) of degree $40$-$60$ percent. The more astonishing result is found in the heat capacity calculations. The ratio of the square root of specific heat and the absolute value of the average energy attains a magnitude between $2.0$ and $3.0$ whereas the expected magnitude is around $0.02$ in thermal equilibrium at any temperature for our system size.

 The above analysis involving the static properties from MC calculations brings out the presence of enormous fluctuations even in the thermal equilibrium. We would now like to present and analyse the results for spin dynamics, as obtained from our MCMD calculations and examine the consequences of fluctuations further.

 The results for the dynamical structure function and the spin auto-correlation function show quite different type of behaviour with regard to the effect of fluctuations, both in the low temperature and in the high temperature regime. In particular, the results for $S({\bf q},\omega)$ exhibit very peculiar and dramatic nature. In contrast, the results for $C(t)$ are more well behaved and do not display any surprising feature. 

 The most interesting and intriguing aspect of the results for spin dynamics is the appearance of the negative values for $S({\bf q},\omega)$ in a very large region of the  $\omega$-space. The details of these striking features are given below. 

 At the temperature corresponding to $16$K, in the $S({\bf q},\omega)$ vs. $\omega$ plot for $\bf q$ along the $<100>$ direction the positive regime dominates over the negative one, particularly in the low $\bf q$ and low $\omega$ regime (see fig. \ref{fig1} (top panel)). In the physically admissible regime $viz.$ the regime with positive values of $\omega$, there are signatures of propagating modes (see fig. \ref{fig1}(middle panel). The propagating modes are characterized by the occcurrences of peaks at finite $\omega$ in the above plots. The quantity $C(t)$ shows a monotonic decay with time (see fig.  \ref{fig1}(bottom panel)).

\begin{figure}
\includegraphics[scale=.3]{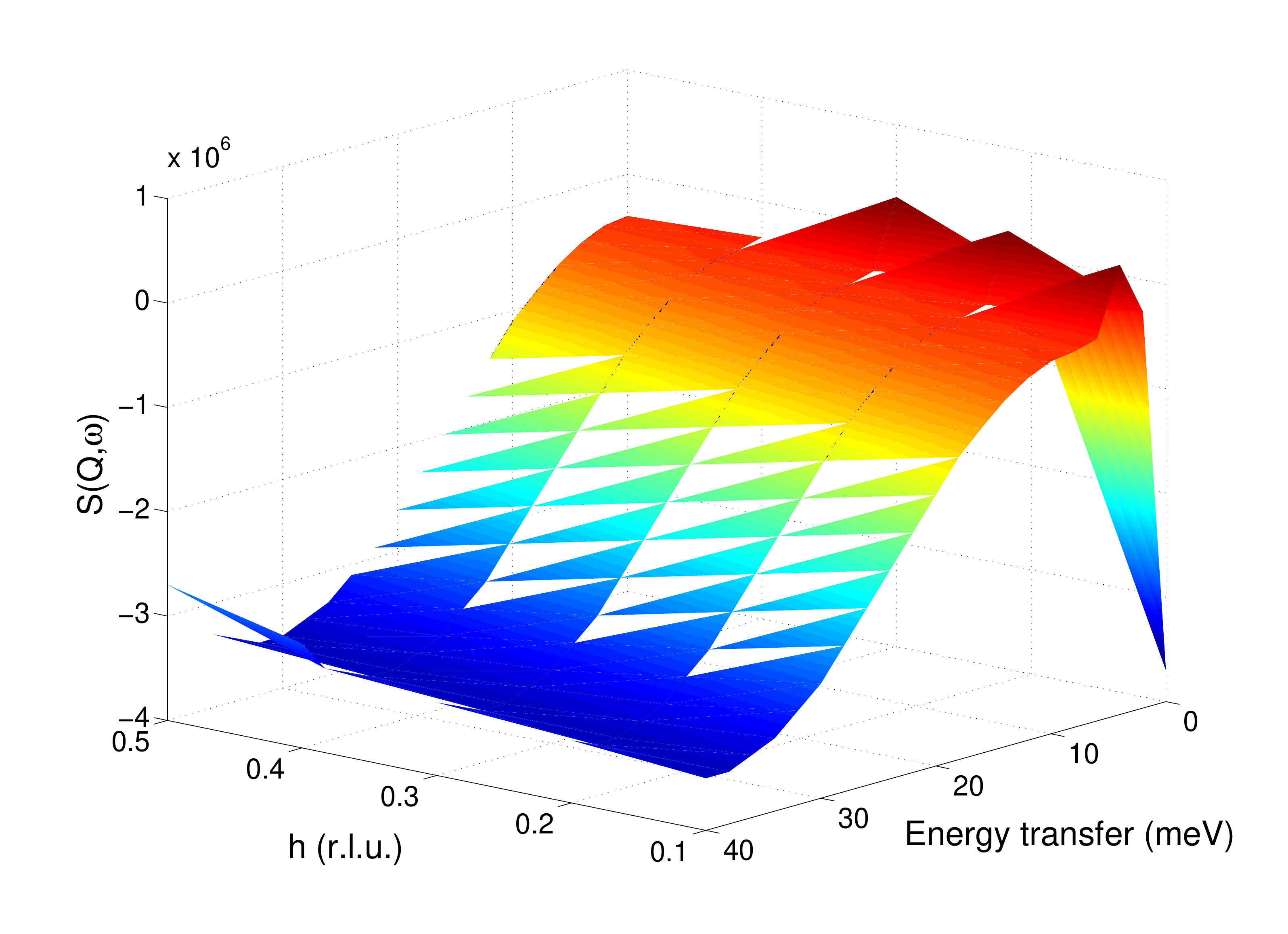}
\caption{Three-dimensional colored plot of high temperature $S({\bf q},\omega)$ vs. $\omega$ for various values of ${\bf q}$ along $<100>$ direction.}
\label{fig3}
\end{figure}

 In the high temperature regime $i.e.$ at $200$K, we observe a complete reversed trend in $S({\bf q},\omega)$. The region with negative values for DSF dominates over that with positive values for almost all the $q$-values along the $<100>$ direction ! Therefore it is conceptually difficult to determine the nature of the collective modes whether they are propagating or diffusive. Furthermore, the shapes of the curves representing $S({\bf q},\omega)$ vs. $\omega$ in the constant $q$-scans display a lot of variety with the occurrences of cusps, plateau and shoulders besides peaks (see fig. \ref{fig2}(top panel) and fig. \ref{fig2}(middle panel)). The SAF shows a monotonic but very sharp decay with time (see fig. \ref{fig2}(bottom panel)). The high temperature DSF vs. $\omega$ plot in the entire $\bf q$ space along the $<100>$ direction, can also be seen in the three-dimensional color graph (see fig. \ref{fig3}).  

 The above two temperature regimes may be characterized as "moderate fluctuation driven" (for the low temperature phase) and "strong fluctuation driven" (for the high temperature phase), on the basis of the analysis of our static results. Our spin dynamics results reaffirm and manifest these effects more clearly. The "unphysical region" observed in the calculations for DSF originates, we believe, from the breakdown of the semi-classical approximation (used in our MCMD) in the presence of the unusually high degree of fluctuations, as we discuss below. 

 It is well known in literature [8] that $S_{QM}({\bf q},\omega)$ (DSF as obtained quantum mechanically) is real and positive whereas $S_{CL}({\bf q},\omega)$ (DSF as obtained classically) can in general be a complex quantity. Furthermore, the dynamical spin fluctuations  and the imaginary part of the dynamical spin susceptibility, as calculated quantum mechanically, can be directly related by the "fluctuation-dissipation" theorem. The dynamical spin fluctuation reduces to the dynamical structure function (DSF) in the paramagnetic phase. Besides, DSF for positive $\omega$ and negative $\omega$ are related by the "detailed balance" condition in thermal equilibrium. These features and properties prompted Windsor [1,9] to propose an ansatz by which one can extract the true quantum mechanical DSF from the one which had been obtained using the nearly classical methodology of MCMD. It should be kept in mind that this ansatz is ideally suited for the paramagnetic phase ; however it may be used even in the ordered phase provided the magnitude of the order parameter is not too large.\\ 

 Mathematically, the ansatz of Windsor is expressed as
\begin{equation}
S_{QM}({\bf q},\omega) ={\frac{2Real{S_{CL}({\bf q}, \omega)}}{1+exp(-{\beta}{\omega}\frac{h}{2\pi})}}
\end{equation}  

 The above prescription takes into account all the above mentioned properties and is essentially based on an ansatz that the classical spin-spin correlation function is equal to the real part of the corresponding quantum function in the time domain. This assumption is meaningful and accurate in the semi-classical limit $i.e.$ genuinely valid in the high $S$ and in the rather low energy (low $\omega$) limit. In fact, it can be shown that as long as the energy under consideration is less than the characteristic energy (${\frac{h}{2\pi}}\omega_c$) corresponding to the "natural time unit" ($t_0$) occurring in our formalism, Windsor's ansatz works well. It may be recalled that ($t_0^{-1}$) is proportional to $J[S(S+1)]^{\frac{1}{2}}$ and this has dimension of energy. Thus it is obvious that this parameter $\omega_c$ increases with $S$, implying a very wide regime for the validity of the semi-classical ansatz in the large spin limit. The microscopic origin of the existence of this characteristic/cutoff energy scale is the genuine semiclassical-like behaviour of the spin dynamics, prevailing in the time scale above the quantum fluctuation (spin flip) time $t_0$. 

 The manifestation of the above property can be seen in our MCMD results on spin models with parameters corresponding to the Rare Earth Chalcogenides like $EuO$ and $EuS$. The cutoff energies turn out to be $4.5$ meV and $0.84$ meV for $EuO$ and $EuS$ respectively. Above these energies the respective DSF's calculated  by our semiclassical MCMD for the above two systems, when plotted in the constant $q$-scans as a function of $\omega$, indeed become negative as can be seen from the corresponding figures [1,2]. Thus we may say that these cutoff energies are highly physical and truly represent the boundary between the semiclassical and the quantum behaviour of the spin dynamics. Furthermore, our theoretical estimates of these energies based on the arguments presented in the earlier paragraph, are also fairly close to the observed ones from the plots [1,2].\\

 Now coming to the case of our present system $viz.$ $La_{1.2}Sr_{1.8}Mn_{2}O_{7}$, we extract the values of the parameter $\omega_c$ from the $S({\bf q},\omega)$ vs. $\omega$ plots obtained from MCMD and also estimate its magnitude from the value of $t_0$. When we compare the two, we see a very large discrepancy! Keeping in mind that $t_0 = {\frac{h}{2{\pi}J_1[S(S+1)]^{\frac{1}{2}}}}$, we get the magnitude of $t_0$ as $6\times 10^{-14}$ seconds. This leads to the magnitude of the critical energy as high as $59$ meV approximately ! Our MCMD results however indicate a negative regime for DSF even below $5$ meV at both $16$K and at $200$K. 

 In order to understand the above peculiarity, we will have to also include the possible effects arising from the thermal fluctuations, seen earlier in the studies of the static properties from MC calculations. The giant fluctuations observed in the thermodynamic properties imply that a wide variance is also expected in the contributions to the various static and dynamic response functions from the different ensemble configurations. The quasi-two dimensionality of the system  is the root cause of these giant thermodynamic fluctuations. Thus the total fluctuations in the system far exceed the amount expected from the pure quantum effects. Furthermore, the existence of these unusually large thermodynamic fluctuations even away from the transition temperature may also cause a departure from the standard fluctuation-dissipation relation, leading to the possible modification of Windsor's Ansatz. As a result, the extracted $S_{QM}({\bf q},\omega)$ from the semiclassical MCMD results (by using the above ansatz), can exhibit a lot of unphysical behaviour and may even become negative !  
 
\begin{figure}
\includegraphics[scale=.45,angle=0]{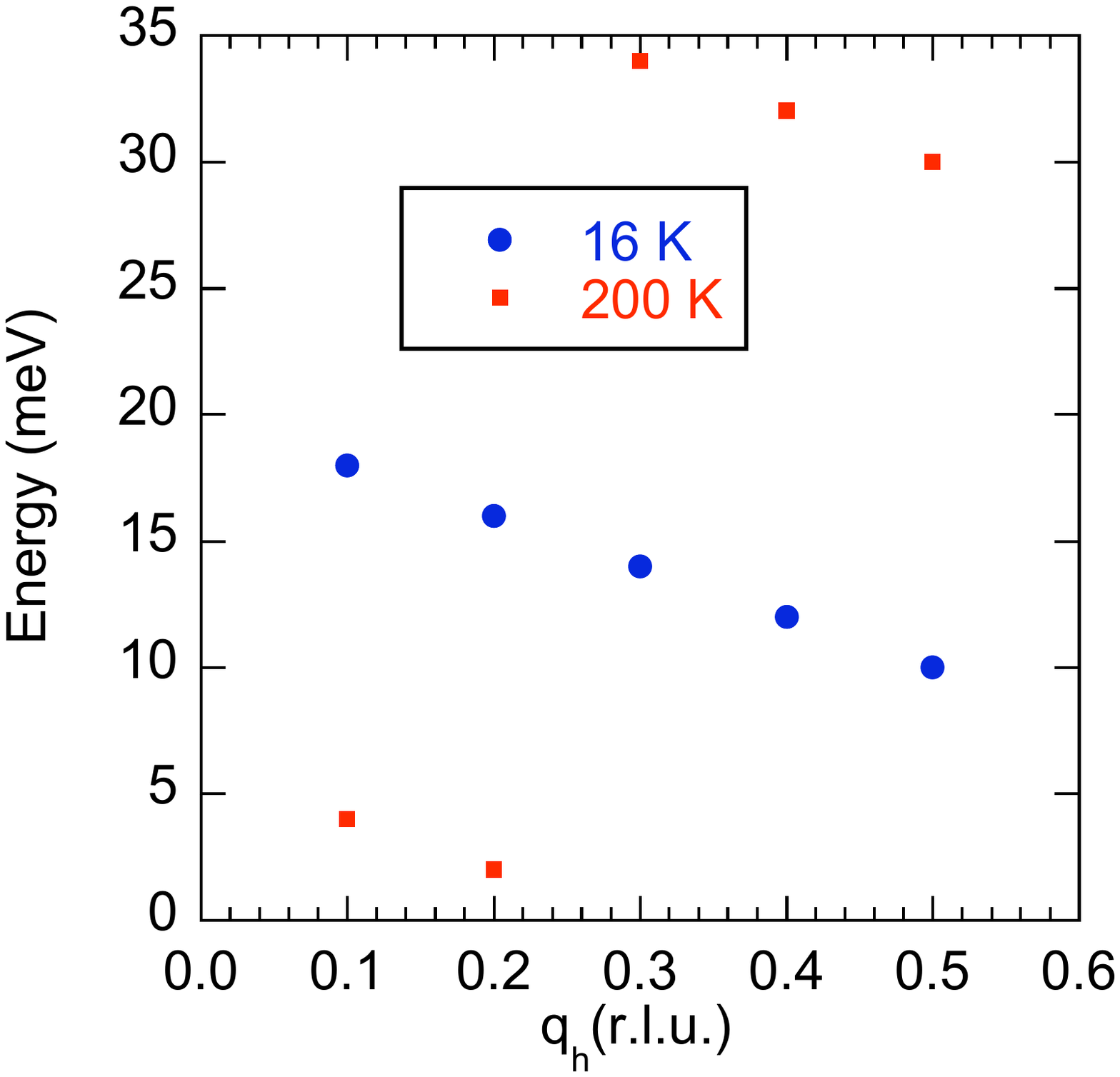}
\caption{(Blue circles) Dispersion $\omega_{\bf q}$ vs. ${\bf q}$ along $<100>$ direction for the low temperature phase $viz.$ at $T=16K$. (Red circles) Dispersion $\omega_{\bf q}$ vs. ${\bf q}$ along $<100>$ direction for the high temperature phase $viz.$ at $T=200K$.}
\label{fig4}
\end{figure}

\section{ Conclusions}
 In this short communication we have tried to present an analysis and a possible interpretation of our calculational results for $S({\bf q},\omega)$ and $C(0,t)$. In particular, we have put forward an explanation for the occurrences of negative values for the extracted $S_{QM}({\bf q},\omega)$ in a large regime of $({\bf q}, \omega)$ space. The calculated spin auto-correlation function from our semiclassical MCMD on the contrary, behaves in an expected way--- exhibiting a faster decay with time at higher temperature. This contrasting behaviour between DSF and SAF arises because the former is a "global property" whereas the latter is a "local property" and that the presence of a high degree of misalignment of spins (seen in the giant fluctuations) affect the former much more than the latter. 

 Furthermore, our quantum mechanical DSF is calculated by using an ansatz which  may not work very well when the thermodynamics is dominated by the very large fluctuations, as is the case here. On the other hand, our SAF is computed directly from the MCMD results and is of semiclassical nature. Therefore, our calculated and extracted estimates of the above two quantities do not obey the theoretically expected consistency relation between them. 

 In the physically admissible regime, the $S({\bf q},\omega)$ vs. $\omega$ plots can be used for studying the nature of excitations and understanding the genuine spin dynamics. The dispersion curves ${\omega}_{\bf q}$ vs. ${\bf q}$ corresponding to the "propagating regime" from our calculations in both the low temperature phase and in the high temperature phase have been presented in the figure \ref{fig4}. These may also be compared with the corresponding experimental results, for the judgement on the rationale of our modelling itself. It is quite striking that the slope of the calculated
         dispersion curve is negative, i.e. opposite to that expected for a 
         full 3D system [3]. In order to understand this, it is to be noted
         that the long range
         order is quite "fragile" in this case and hence the collective
        propagating modes can show quite different behaviour in its dispersion
        from that in a pure 3D system. In particular, the propagating modes
        become more ill-defined with increase in the magnitude of "q" because
        of "disorder" present in the short length scales within the spin
        configurations in the ordered phase, causing a softening of the modes as
        well.  In addition, the inadequacy of the
        semi-classical treatment can cause more deviation from the expected
        behaviour in $S(q,\omega)$ results, as obtained by MCMD calculations. These
        lead to the drastically different slope of the dispersion curve of 
        the propagating modes in the ordered phase in our MCMD calculations,
        from that for a full 3D system.
        In the paramagnetic phase, the behaviour
        is even more complicated as there is a competition between the "short
        range ordering" expected for a full 3D system and "fragility" inherent
        within the spin configurations in a quasi-2D system. This causes the 
        dispersion curve to first have a rise and then a fall. 

 Before concluding, it should be emphasized that the presence of this high degree of thermodynamic fluctuations in the static properties seen from the MC calculations after the convergence, is not at all inconsistent with the attainment of  "steady state" for the spin system. This state however does not always ensure complete "thermal equilibrium" for the system, as is understood in the conventional way. Nevertheless, even in the complete thermal equilibrium condition as is well known, the magnitude of fluctuations can very well exceed the usual expected estimate $viz.$ ${\frac{1}{N^{0.5}}}$ under certain conditions, like in the vicinity of phase transition for example. In our case however, the low dimensionality $viz.$ the quasi-two dimensional nature of the system (lattice) appears to be the most likely origin of these enormous amount of thermodynamic fluctuations occurring both in the ordered as well as in the disordered phase.\\

\end{document}